\documentclass[times, twoside]{zHenriquesLab-StyleBioRxiv-Submission}

\usepackage{blindtext}
\usepackage{graphicx}
\usepackage{dblfloatfix} 
\usepackage{lineno}
\usepackage{comment}
\usepackage[normalem]{ulem}
\usepackage{xcolor, colortbl}
\usepackage{MnSymbol}

\renewcommand{\thesubsection}{\thesection\Alph{subsection}}

\renewcommand{\footerfont}{\normalfont\sffamily\fontsize{8}{9}\selectfont}

\leadauthor{Cornwall} 

\begin{document}

\title{\textcolor{black}{Synthetic Light-in-Flight}}

\shorttitle{Synthetic Light-in-Flight}

\author[1,*]{Patrick Cornwall}
\author[2]{Manuel Ballester}
\author[1]{Stefan Forschner}
\author[1]{\\Muralidhar Madabhushi Balaji}
\author[2,3]{Aggelos Katsaggelos}
\author[1,2,3,*]{Florian Willomitzer}

\affil[1]{Wyant College of Optical Sciences, University of Arizona, Tucson, AZ, 85721, USA.}
\affil[2]{Department of Computer Science, Northwestern University, Evanston, IL 60208, USA.}
\affil[3]{Department of Electrical and Computer Engineering, Northwestern University, Evanston, IL 60208, USA.}

\affil[*]{patrickcornwall@arizona.edu, fwillomitzer@arizona.edu}

\maketitle

\vspace{-3mm}

\section*{Abstract} 
\textit{Light-in-flight (LiF) measurements enable the visualization of light paths through arbitrary, volumetric scenes, making light-matter interactions at ultrafast timescales visible. Traditionally, LiF measurements require specialized equipment, such as ultrashort pulse light sources and high-speed electronics, often limited by low spatial resolution. Herein, we introduce a novel computational approach,``Synthetic Light-in-Flight'' (SLiF), that overcomes these constraints by relying solely on tunable, continuous wave (CW) lasers and off-the-shelf CMOS cameras. From multiple CW scene measurements at different optical wavelengths, we create multiple ``synthetic fields,'' each at a ``synthetic wavelength,'' which is the beat wave of two respective optical waves. These synthetic fields are robust to speckle and environmental fluctuations, enabling us to combine multiple synthetic fields into a ``synthetic light pulse'' that sections the volumetric scene at much lower instantaneous peak illumination power than a comparable physical light pulse. We experimentally demonstrate the generation of synthetic pulses with  $ 1ps$-scale width and  show that their} complex synthetic pulse fields can be freely manipulated in the computer after their acquisition, allowing for spatial and temporal shaping of different sets of pulses from the same set of measurements to maximize the decoded information output for each scene. Finally, we show that the recovered time-of-flight information can be used to characterize physical scene properties, such as depth and refractive indices.

\section{Introduction}
\label{sec:introduction}
Although many years have passed since light-in-flight (LiF) measurements were first demonstrated \cite{abramson.1978, abramson.1983, Häusler.1996, inoue.2023, faccio.2018}, visualizations of a light pulse traveling through a volumetric scene and interacting with embedded objects have still not lost their ``magical effect'' on the observer. Today, light-in-flight imaging (i.e., the ability to resolve light itself in space and time at high frame rates as it propagates through a volumetric scene) has developed into a powerful method to observe light-related ultrafast phenomena, which in turn enables numerous potential imaging applications in biomedical imaging, industrial inspection, or virtual reality. Examples include observing light propagation through heterogeneous material or tissue \cite{Häusler.1996, liang.2018, satat.2016, boccolini.2017}, characterization of manufactured components \cite{wilson.2017, liu.2024}, and the development of innovative viewpoint rendering techniques \cite{malik.2024, malik.2025}. 

State-of-the-art LiF techniques generally capture LiF videos by launching ultra-short light pulses into a medium or volumetric scene and recording the response with a high-speed detector such as a single-photon avalanche diode (SPAD) or a streak camera \cite{faccio.2018, velten.2013, pifferi.2016, gariepy.2015, heshmat.2014, gao.2014, wang.2020, nousias.2025}.  As an example, the iconic ``soda bottle video''~\cite{velten.2013} was captured using this kind of approach, and related LiF concepts led to important follow-up ideas in research fields such as non-line-of-sight imaging or imaging through scattering media \cite{otoole.2018, faccio.2020, lindell.2019, liu.2019, velten.2012, lindell.2020, satat.2016}.

In this paper, we present a novel computational approach to LiF imaging that completely eliminates the need for high-speed imaging equipment and light pulses with high instantaneous intensity. Instead, our technique uses ordinary high-resolution CMOS detectors (3648 $\times$ 3648 pixels) and (tunable) continuous-wave (CW) laser sources. Our approach draws inspiration from a variety of interferometric and holographic LiF approaches that have been proposed over the years \cite{arons.1995, Häusler.1996, inoue.2023, shih.1999, marron.1992}, but specifically builds upon the concept of ``synthetic wavelength imaging''  (SWI), which has been utilized for various applications in 3D imaging of rough surfaces, non-line-of-sight imaging, and imaging through scattering media and optical fibers \cite{fercher.1985, dandliker.1988, willomitzer.2021, willomitzer.2019, willomitzer.2024, ballester.2024, rangarajan.2019, li.2021, li.2018, kotwal.2023, groger.2023, degroot.1992, Cornwall.2023, Forschner.2024, kassem2025intensity, balaji.2026}: When coherent light is reflected off a rough surface or transmitted through a scattering medium, the phase randomization in the arising speckle field hinders the extraction of distance or time-of-flight information from an interferometric measurement. SWI addresses this problem by exploiting spectral correlations in speckle fields captured full-field at slightly different wavelengths $\lambda_1$ and $\lambda_2$. Combining these two speckle fields $E(\lambda_1)$, $E(\lambda_2)$ creates a high frequency carrier wave with a low frequency beat note. The wavelength of this beat wave can be described by the ``synthetic wavelength''   $\Lambda = \lambda_1 \cdot \lambda_2 / |\lambda_1 - \lambda_2|$ \cite{willomitzer.2024, willomitzer.2021}.  For closely spaced $\lambda_1$ and $\lambda_2$, which results in a large $\Lambda$,  the computationally generated ``synthetic field'' $E(\Lambda)$ is largely robust to the phase perturbations in an optical speckle field, and hence can be used for interferometric phase measurements of light transmitted through a scattering medium or reflected off a rough surface. The phase evaluation is then performed purely computationally at the much larger synthetic wavelength $\Lambda$ \cite{willomitzer.2024, willomitzer.2021}.

In SLiF, multiple of these synthetic fields are superimposed computationally to form a ``synthteic pulse'' that virtually sections the respective scenes.  This procedure offers several practical advantages that distinguish it from conventional pulsed approaches: The synthetic pulse can be computationally generated  after acquisition, and its spatial and temporal shape  can be flexibly tailored to specific scenarios without modifying the laser illumination or detection hardware. This enables a multitude of flexible pulse-like illumination patterns that can be freely explored computationally post acquisition. Moreover, because the effective temporal structure is encoded using continuous-wave illumination rather than ultrashort laser pulses, SLiF offers the unique possibility of recovering pulse-like, time-resolved information while avoiding the high instantaneous peak intensities associated with conventional pulsed illumination. Beyond the experiments shown in this paper, this may be advantageous in nonlinear, scattering, or resonant photonic systems where pulsed excitation can alter the system under investigation, for example through intensity-dependent refractive-index changes, nonlinear scattering processes, or resonance shifts.

In the following sections of this paper, we describe our ``Synthetic LiF''  (SLiF) approach and provide proof-of-principle experimental results, serving as a first-ever demonstration of pulsed SLiF measurements in a variety of volumetric 3D  scenes with varying refractive indices, and scattering media with embedded foreign bodies. Additionally, we demonstrate for the first time how SLiF pulses can be computationally shaped and manipulated in time and space \textit{after} their acquisition to highlight various features and maximize the decoded information output for each measured scene.

\section{From Synthetic Waves to Synthetic Pulses}
\label{sec:wavestopulses}

\begin{figure*}[t!]
\centering
\includegraphics[width=\linewidth]{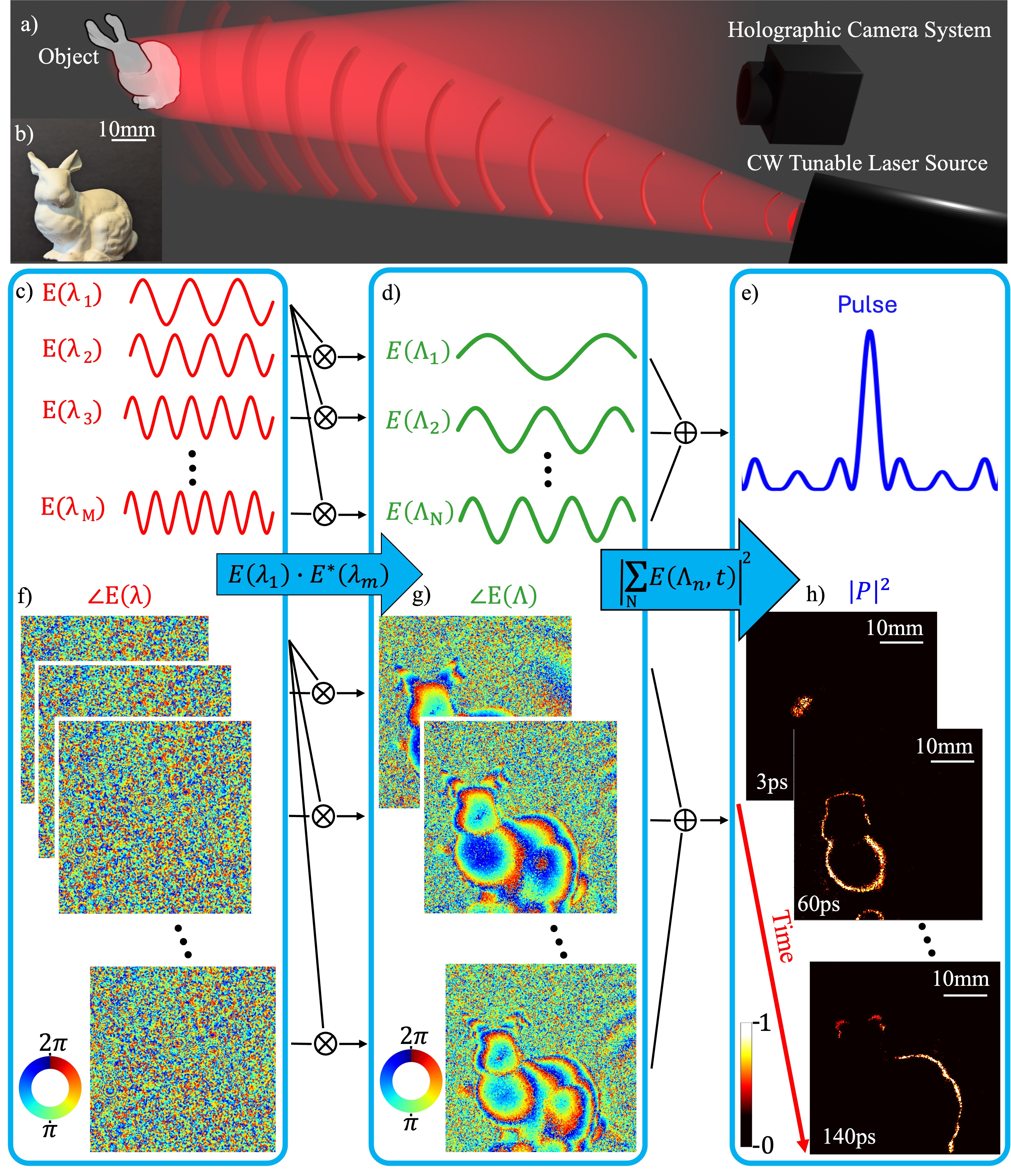}
\vspace{-9mm}
\caption{\textbf{Synthetic Light-in-Flight (SLiF): Generation of synthetic pulses:} a) Schematic of measurement setup: A scene is illuminated at multiple wavelengths and a holographic camera \cite{ballester.2024} captures the field information of each back-scattered optical field. b) Image of measured object. c-e) The set of captured optical fields (c) at different wavelengths  $\lambda_{1}, ...,  \lambda_{M}$ is used to create a set of  synthetic fields (d) at different synthetic wavelengths $\Lambda_{1}, ...,  \Lambda_{N}$  via computational pairwise mixing. The synthetic fields are computationally phase-aligned and superimposed to create  a synthetic pulse train (e). The pulse becomes more well-defined, the more synthetic fields are added. f-h)~Example images for: Speckled phase maps $\phi(\lambda_m)$ of captured optical fields (f). Phase maps $\phi(\Lambda_n)$  of calculated synthetic fields (g). Squared amplitude of assembled synthetic pulse $|P|^2$, shown at different time stamps between 0ps and 140ps. It can be seen that the assembled synthetic pulse precisely sections the object surface as it advances through the scene (see video \href{https://drive.google.com/drive/folders/1rS9Itz3QuB3RZksTBMV8XeLqz7HhctGi?usp=drive_link}{here} \cite{videos.2024}). The experimentally evaluated pulse FWHM for this measurement is $9.88\ \mathrm{ps}$ or $2.96\ \mathrm{mm}$.}
\vspace{-1mm}
\label{fig:Fig_1}
\end{figure*}

Our procedure for synthetic pulse generation starts with the synthetic wave field acquisition of the scene or object of interest. As described in \cite{willomitzer.2024, willomitzer.2021}, one method for generating an unspeckled synthetic wave field $E(\Lambda)$ is to computationally mix the two
 optical speckle fields $E(\lambda_{1})$ and $E(\lambda_2)$ that can be captured full-field without scanning (see Fig.~\ref{fig:Fig_1}):

\vspace{-6mm}

\begin{align}
    E(\Lambda) &= E(\lambda_{2})\cdot E^*(\lambda_1) \notag \\
    &= A_{\lambda_{2}} A_{\lambda_1} \cdot e^{i(\phi(\lambda_{2}) - \phi(\lambda_1))} 
    = A_\Lambda \cdot e^{i\phi(\Lambda)} 
\label{eq:SynthField}
\end{align}
 
\vspace{-2mm}

In previous work \cite{ballester.2024}, our team has demonstrated motion-robust, holographic acquisition of $E(\Lambda)$ in single-shot, by simultaneously capturing the two complex speckle fields, $E(\lambda_{1})$ and $E(\lambda_2)$, in one camera image of an ordinary CMOS camera. For the purpose of this paper, we use the same off-axis holographic acquisition scheme \cite{ballester.2024} and camera system (in the following referred to as ``holographic camera''), but capture $E(\lambda_{1})$ and $E(\lambda_2)$ individually in a sequential fashion (each captured respectively in single-shot), as this acquisition scheme delivers slightly better noise characteristics. In this case, their beat wave $E(\Lambda)$ (shown in Fig.~\ref{fig:Fig_1}d) has never existed physically and exists only on the computer. Nevertheless, the complex-valued synthetic field $E(\Lambda)$ possesses characteristics largely similar to those of a conventional electromagnetic field at wavelength~$\Lambda$. In particular, if $\lambda_1$  and $\lambda_2$ are closely spaced, the phase of the synthetic field $\phi(\Lambda) = \phi(\lambda_{2})-\phi(\lambda_1)$ is unperturbed by microscopic path length variations introduced when light is reflected off an optically rough surface or propagates through a scattering medium~\cite{willomitzer.2024, willomitzer.2021}. This is because the synthetic wavelength $\Lambda$ can be chosen orders of magnitude larger than the individual optical wavelengths (see Fig.~\ref{fig:Fig_1}d). At the same time, the same synthetic wavelength~$\Lambda$ can be synthesized in different wavebands (i.e., VIS or IR) allowing optimal transmission or reflection properties for a respective application. For carrier wavelengths in VIS-NIR, ordinary CMOS cameras can be used for detection, which typically display high sensitivity and superior pixel resolution.

One key contribution of this paper lies in the realization of how the concept of synthetic waves can be used for pulsed LiF measurements. The proposed procedure exploits another striking analogy between synthetic waves and optical waves: For pulse generation in the optical domain (e.g., in a pulsed laser), light fields at multiple wavelengths are synced in phase and simultaneously emitted so that their superposition in space forms a light pulse. To generate and capture a ``synthetic pulse'', we computationally perform the ``synthetic wave equivalent''  of optical pulse generation: The basic idea is to capture multiple synthetic fields $E(\Lambda_n)$ ($n=1,...,N$; see Fig.~\ref{fig:Fig_1}d), each generated from a pair of captured optical fields $E(\lambda_1)$ and $E(\lambda_m)$ ($m=1,...,M$, see Fig.~\ref{fig:Fig_1}c). After their acquisition, these synthetic fields are eventually computationally superimposed to form a synthetic pulse $P$ \cite{Cornwall.2023} (see Fig. \ref{fig:Fig_1}e).

In real-world experiments, drift and other low frequency perturbations prevent  the phases of the captured fields $E(\Lambda_{n})$ from aligning after their acquisition and an alignment procedure must be performed prior to their superposition. Notably, the unique properties of synthetic fields (in particular, their smooth, unspeckled phase front) allow us to perform this alignment completely computationally after acquisition: As shown in Eq.~\ref{eq:SynthField}, each generated synthetic field is assigned a synthetic amplitude $A_{\Lambda_n}$ and synthetic phase $\phi(\Lambda_n)$, both of which can be freely manipulated in the computer after acquisition. For example, we can add a global phase offset to all phase values in the 2D arrays of every captured synthetic field~$E(\Lambda_{n})$. To computationally align the phase values in all captured fields, we observe the captured synthetic phases $\phi(\Lambda_n, x_o,y_o)$ at an arbitrary scene point $(x_o,y_o)$. Eventually, we add a global phase offset $\Delta \phi_n$ to each individual synthetic field (new phase: $\tilde{\phi}(\Lambda_n) = \phi (\Lambda_n) + \Delta \phi_n$), satisfying the condition of Eq.~\ref{eq:phasesync} that all phase maps for all  synthetic fields arrive at the same phase value in point $(x_o,y_o)$:
\vspace{-1mm}
\begin{equation}
 \tilde{\phi}(\Lambda_{1}, x_o,y_o) = \tilde{\phi}(\Lambda_{2}, x_o,y_o) = ... = \tilde{\phi}(\Lambda_{N}, x_o,y_o)
\label{eq:phasesync}
\end{equation}

\vspace{-1mm}

We emphasize again that this computational alignment is not possible with scattered fields at the (optical) carrier wavelengths, as their phase values are randomized due to speckle. 
After phase alignment,  we additionally replace the (speckled) amplitude $A_{\Lambda_n}$ of each synthetic field with an array of ones ($\tilde{A}_{\Lambda_n} = 1$) to reduce speckle artifacts in our final pulse reconstruction (see discussion in sec. \ref{subsec:limitations} ). With each computationally manipulated synthetic field \mbox{$\tilde{E}(\Lambda_n) = \tilde{A}_{\Lambda_n} \cdot e^{i\tilde{\phi}(\Lambda_n)}$}, the synthetic pulse is then assembled via

\vspace{-5mm}
\begin{equation}
    P(\Lambda_1, \ldots, \Lambda_N) = \tilde{E}(\Lambda_{1}) + \tilde{E}(\Lambda_{2}) + \cdots + \tilde{E}(\Lambda_{N})
\label{eq:addingfields}
\end{equation}
\vspace{-5mm}

Due to the computational phase alignment at scene point $(x_o,y_o)$,  calculation of the intensity term $|P|^2$  produces a ``synthetic pulse maximum''  at  $(x_o,y_o)$, \textit{and all scene points that observe the same optical path length distance} (see Fig.~\ref{fig:Fig_1}e). This results in a pulse-front which is then computationally advanced over the scene by simply advancing the time variable in the synthetic phase term of each synthetic field $\tilde{E}(\Lambda_n)$. We emphasize that the alignment procedure discussed above only assumes global phase changes (e.g., due to laser drift) in between the subsequently captured optical fields $E(\lambda_m)$ that transfer to the synthetic fields $E(\Lambda_{n})$. Spatially varying phase changes, such as in strong turbulence, would not yield a clear pulse front as the as the spatial phase offset  distribution changes across the set of captured optical field. In these scenarios, it would become necessary to capture each synthetic field in single-shot, as our team has demonstrated in \cite{ballester.2024, balaji.2026}. Since most synthetic wavelengths used in this work are orders of magnitude larger than the optical wavelengths the residual perturbations would need to spatially vary on the scale of the synthetic wavelength, rather than the optical wavelength, to produce significant errors.

After the synthetic pulse has been assembled and propagated by advancing the respective time variables, we can create a ``Synthetic Light-in-Flight'' video that reveals how the synthetic pulse has traveled through the volumetric scene (see next sections and Figs.~\ref{fig:Fig_2}, \ref{fig:Fig_3}, \ref{fig:Fig_4}, \ref{fig:Fig_6}, and \ref{fig:Fig_8}). Here, the computational pulse advancement can be evaluated at arbitrary small user-defined temporal sampling intervals, allowing the resulting SLiF video to be sampled at high temporal rate. This is distinct from LiF approaches using ``fast cameras'' \cite{faccio.2018, velten.2013, gao.2014, gariepy.2015}, where limiting hardware parameters such as detector bin, gate width, or frame rate influence the temporal sampling of the pulse reconstruction and can affect the temporal resolution. In SLiF, the temporal resolution (which is directly proportional to the depth resolution of the measured scenes) is solely defined by the width of the synthetic pulse, which can be quantified by evaluating its full-width-at-half-maximum (FWHM). In theory, this width is proportional to the inverse of the total captured optical wavelength range $\Delta\lambda = |\lambda_1 - \lambda_M|$ (or ``spectral width'') (see Sec.\ref{subsec:3dimaging}). This relation is also observed in other interferometric LiF methods described in the literature \cite{abramson.1978, inoue.2023, Häusler.1996}. The expressions for the FWHM for various pulse shapes as a function of their spectral bandwidth are well established \cite{weiner.1988, diels.2006}. For the rectangular spectral window and dense spectral sampling used for most experiments in this paper, the corresponding relationship can be approximated by $\Delta t_{\mathrm{FWHM}}~\approx~\frac{0.886 \, \lambda_0^2}{c \, \Delta\lambda}$ (we also refer to our discussion in sec.~\ref{subsec:limitations} and supplementary sec. 1 that considers the influence of noise on our measurements). 
Another important factor is the pulse period, which determines the unambiguous measurement range, and is mainly dependent on the inverse of the smallest spacing between captured individual optical wavelengths, $ |\lambda_{m+1} - \lambda_{m}|$. Both quantities can be equivalently expressed  using synthetic wavelength spacings and are ultimately constrained by the wavelength sampling as defined, e.g., by the tuning characteristics of the used laser (see Sec.~\ref{subsec:3dimaging}).

\section{Synthetic Pulse Propagation and Imaging}

\begin{figure}[b!]
\centering
\vspace{-6mm}
\includegraphics[width=\linewidth]{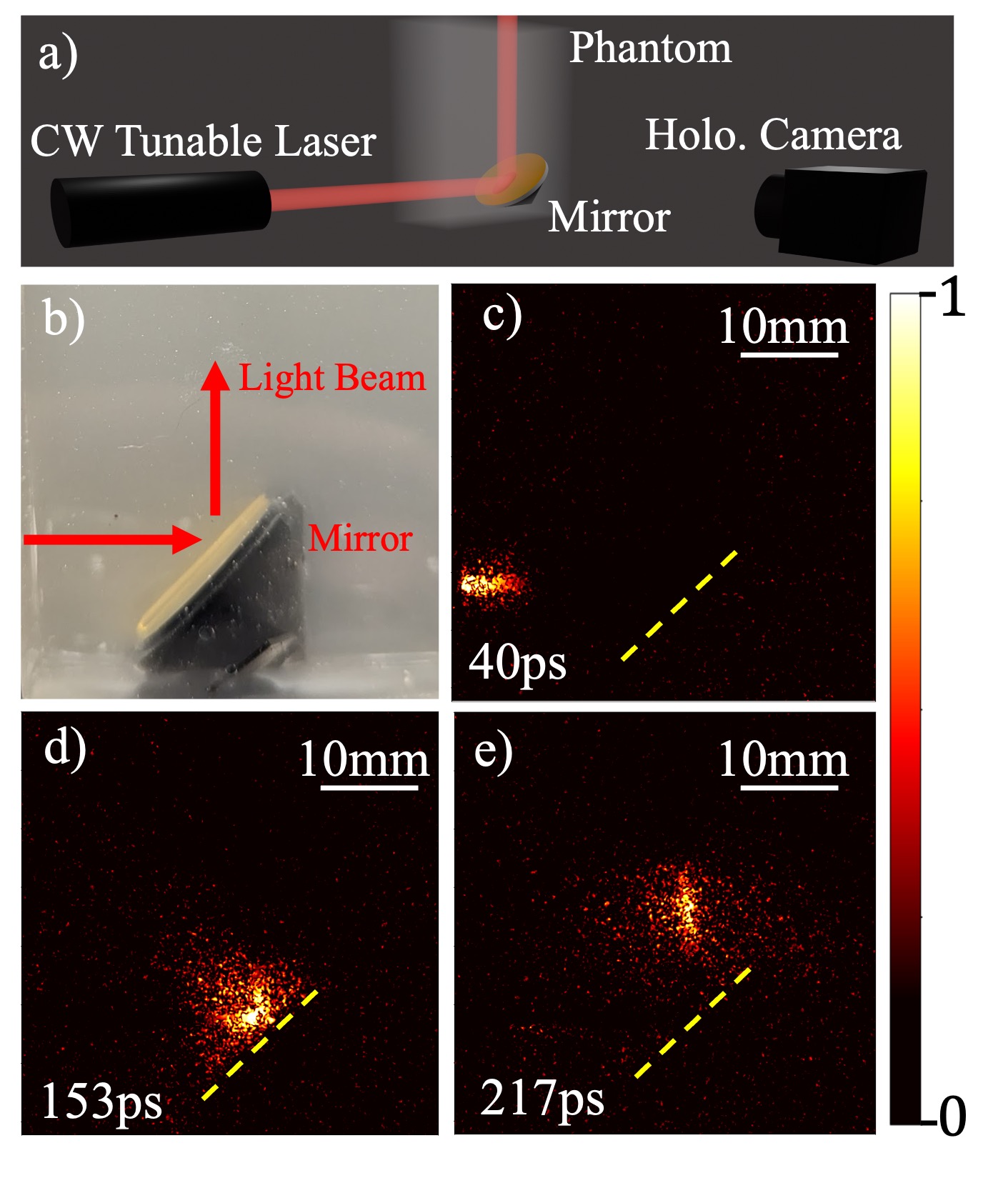}
\vspace{-10mm}
\caption{\textbf{Synthetic Pulse Propagation Through a Volume (see videos \href{https://drive.google.com/drive/folders/1rS9Itz3QuB3RZksTBMV8XeLqz7HhctGi?usp=drive_link}{here} \cite{videos.2024}):}
a) Setup schematic: A collimated laser beam is emitted into a lightly scattering medium and reflected off a $45^\circ$ angled mirror. The holographic camera observes the scene normal to the light propagation plane.  
b) Image of the scene and embedded mirror from the camera's perspective and schematic light path.
c) Computed synthetic pulse entering the scattering medium at $40\ \mathrm{ps}$.
d) The pulse is computationally advanced and reflected off the mirror (dashed yellow) at around $153\ \mathrm{ps}$.
e)~The pulse continues propagation in reflected direction.}
\label{fig:Fig_2}
\end{figure}

In this section, we experimentally demonstrate the ability of our approach to generate LiF measurements that can be used to probe specific scenes or volumes and to obtain volumetric shape information of objects under test.  For the results presented in Figures \ref{fig:Fig_1} to \ref{fig:Fig_4} and \ref{fig:Fig_6} to \ref{fig:Fig_8}, we captured optical fields at $ M=41$ discrete, equally spaced optical wavelengths located within our full laser tuning range from $854.20\ \mathrm{nm}$ to $856.05\ \mathrm{nm}$. This enables the synthesis of $ N = 40$ synthetic fields at distinct synthetic wavelengths. We note that the full tuning range is not used for all presented experiments and that the actual wavelength combinations used vary depending on the depth of the scene being imaged. Shallow scenes, which require a smaller unambiguous depth range, were recorded using larger separations between wavelengths and exploit most of our available tuning range, whereas deeper scenes required finer spacing. 

A measurement taken with a large optical wavelength range (see measurements  performed at $\Delta \lambda = 2.41 nm$ in Fig.~\ref{fig:Fig_5} and $\Delta \lambda = 1.85 nm$ in Fig.~\ref{fig:Fig_6})  yield  experimentally determined pulse FWHM of around $1\ \mathrm{ps}$. This resolution can be improved by increasing the available spectral range. We will conclude this section with a discussion of the involved performance limitations.

\subsection{Synthetic Pulse Propagation Through a Volumetric Scene}
\label{subsec:volumetric}

The measurement shown in Fig.~\ref{fig:Fig_1} depicts the propagation of a synthetic pulse-front across a 3D-printed bunny figure (Fig.~\ref{fig:Fig_1}b). The scene is flood-illuminated at the selected optical  wavelengths $\lambda_m$ from a distant fiber tip, meaning the resulting synthetic pulse-front propagating towards the  bunny figure is not perfectly planar (i.e., parallel to the x-y-plane), and slightly spherical. Figure~\ref{fig:Fig_1}h shows the advancement of the assembled pulse-front over the bunny surface at selected timestamps ($3\ \mathrm{ps}$, $60\ \mathrm{ps}$, $140\ \mathrm{ps}$). The synthetic pulse-front can be seen traveling through the scene and sectioning the bunny surface at different depth values. We encourage the reader to view the rendered SLiF videos \href{https://drive.google.com/drive/folders/1rS9Itz3QuB3RZksTBMV8XeLqz7HhctGi?usp=drive_link}{here}~\cite{videos.2024}.  For the used optical tuning range  ($854.20\ \mathrm{nm}$ to $854.43\ \mathrm{nm}$ in this measurement), the experimentally evaluated FWHM of the visualized synthetic pulse $|P|^2$ is  $9.8\ \mathrm{ps}$, which corresponds to $2.96\ \mathrm{mm}$.

The experiment shown in Fig.~\ref{fig:Fig_2}a-e depicts the propagation of a synthetic pulse packet through a lightly scattering material (necessary to image the packet) and the  interaction with a reflecting mirror. Here,  a tightly collimated laser beam is launched perpendicular to the optical axis towards a $45^\circ$ angled mirror (see Fig.~\ref{fig:Fig_2}). It can be seen that the synthetic pulse propagates through the medium, hits the mirror, and continues in a different direction as dictated by the law of reflection (see SLiF video \href{https://drive.google.com/drive/folders/1rS9Itz3QuB3RZksTBMV8XeLqz7HhctGi?usp=drive_link}{here} \cite{videos.2024}).

\vspace{-2mm}
\subsection{Synthetic Pulse Propagation through Scattering Media}
\label{subsec:scattering}

\begin{figure}[b!]
\centering
\vspace{-6mm}
\includegraphics[width=\linewidth]{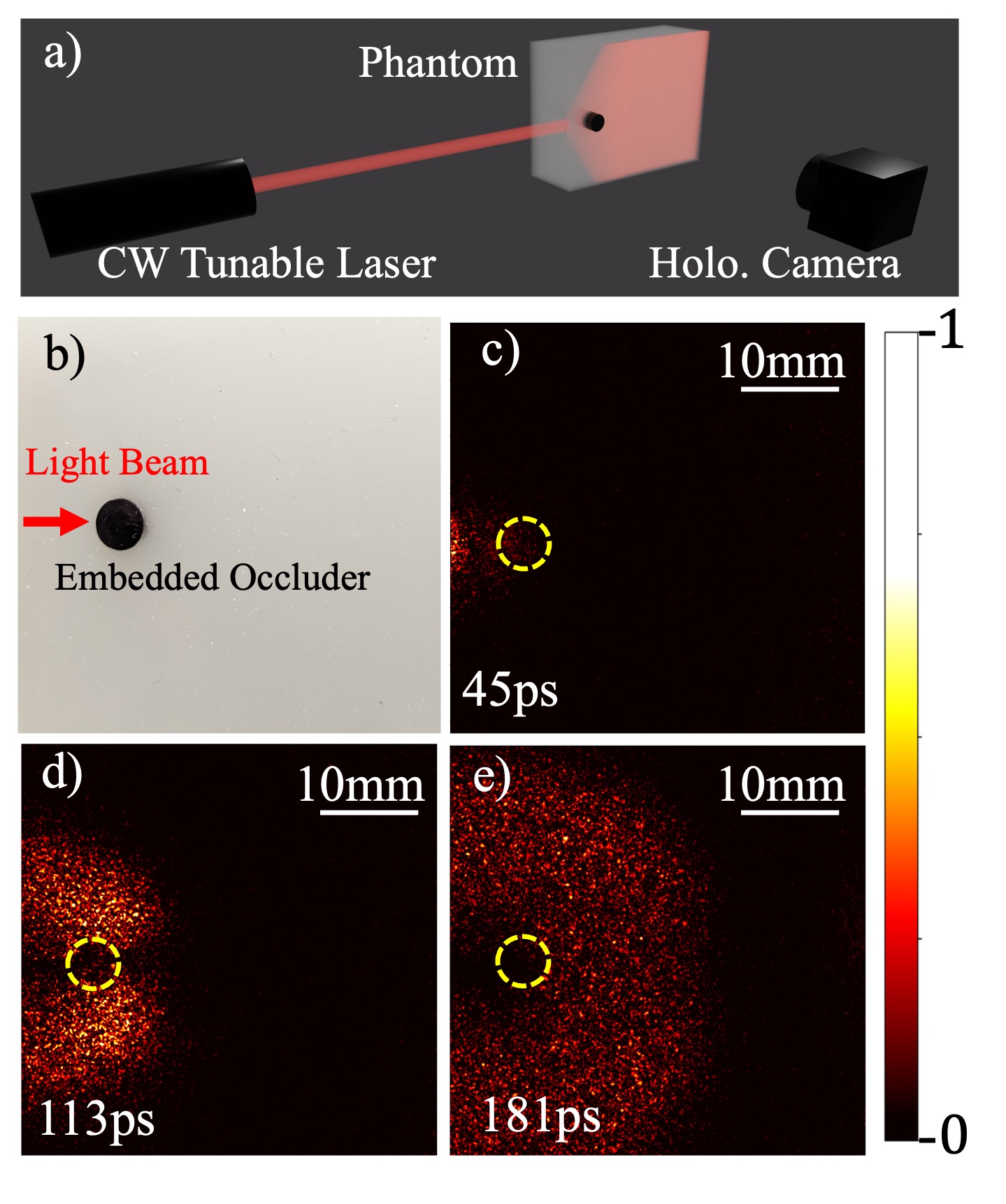}
\vspace{-10mm}
\caption{\textbf{Synthetic Pulse Propagation Through a Heavily Scattering Volumetric Medium (see videos \href{https://drive.google.com/drive/folders/1rS9Itz3QuB3RZksTBMV8XeLqz7HhctGi?usp=drive_link}{here} \cite{videos.2024}):}
a) Setup schematic: A collimated laser beam is emitted into strongly scattering medium with an embedded occluder. The camera observes the scene at a $90^\circ$ angle.
b) Image of the scattering medium with embedded occluder.
c) The computational pulse entering the scattering medium at $45\ \mathrm{ps}$.
d) The pulse is computationally advanced and begins to spread, creating a photon horizon, blocked by the occluder (yellow dashed circle) at $113\ \mathrm{ps}$.
e) Photon horizon traveling around the occluder back into a circular propagation path as it passes across the medium.}
\label{fig:Fig_3}
\end{figure}

The experiment shown in Fig.~\ref{fig:Fig_3}a-e showcases how the SLiF framework can be used to visualize the ``photon horizon'', seen when a light pulse interacts with a strongly scattering medium containing an embedded occluder. Again, a tightly collimated beam is launched perpendicular to the optical axis into the scattering medium and the scattered light is viewed from an orthogonal orientation relative to the propagation direction. While the synthetic pulse advances and spreads through the scatterer, one can observe the propagating photon horizon, as it is blocked by the occluder and eventually scatters back into its geometrical shadow - similar to earlier demonstrations  of LiF through scattering media in \cite{Häusler.1996}. 

\begin{figure}[b!]
\centering
\vspace{-7mm}
\includegraphics[width=\linewidth]{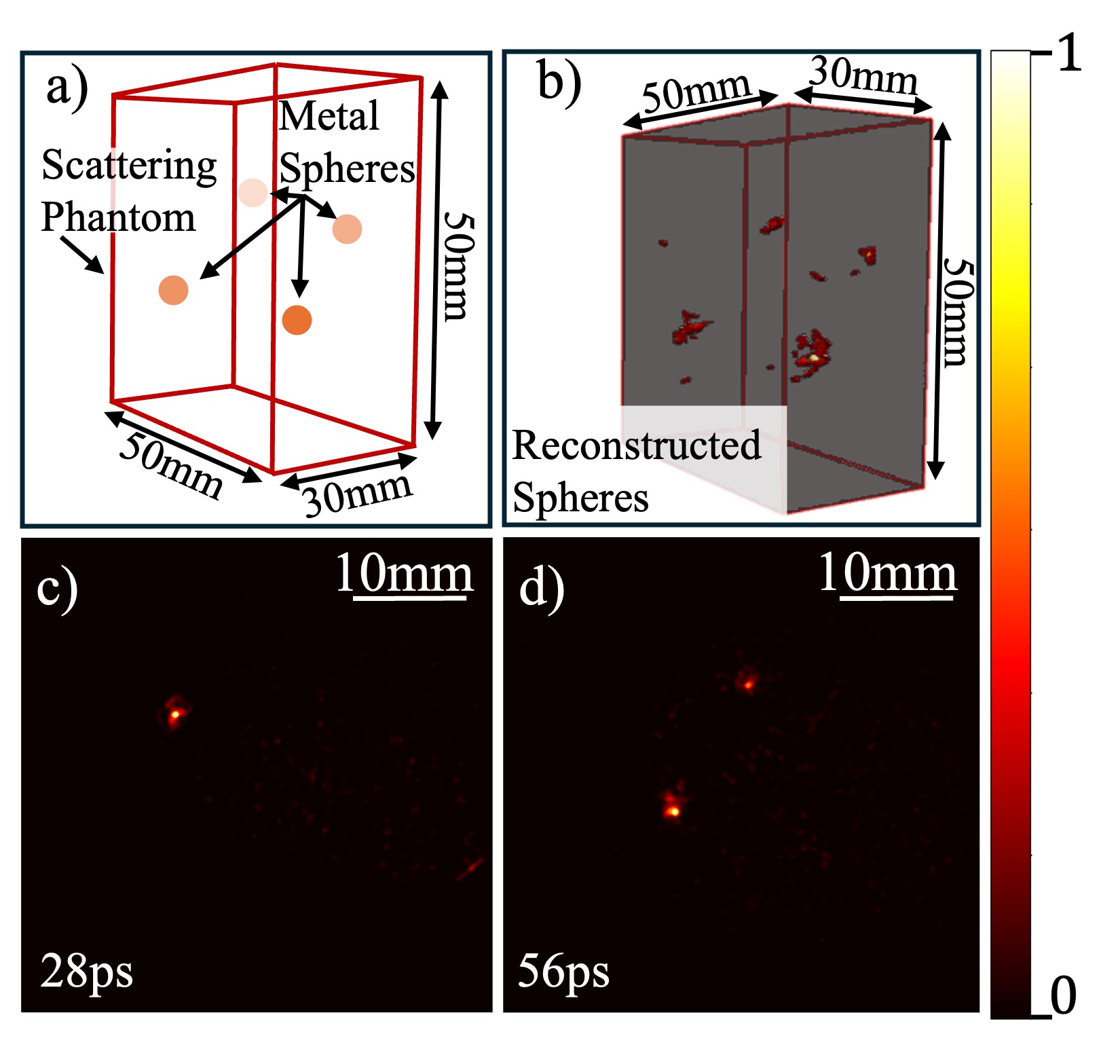}
\vspace{-10mm}
\caption{\textbf{Synthetic Pulse Propagation through a Scattering Phantom with Embedded Foreign Bodies:}
a) Schematic of object: Copper spheres with $4.5\ \mathrm{mm}$ diameter are embedded in a scattering phantom  at various depths ($5\ \mathrm{mm}$–$20\ \mathrm{mm}$).
b) Volumetric reconstruction of sphere locations from time-of-flight data of the SLiF measurement.
c) A time slice highlighting the intensity of the reflection arriving from a single sphere at $28\ \mathrm{ps}$.
d) Second time slice captured at $56\ \mathrm{ps}$ highlighting reflections from two metal spheres.}
\label{fig:Fig_4}
\end{figure}

In another experiment demonstrated in Fig.~\ref{fig:Fig_4}, we use a synthetic pulse to virtually ``slice'' a volumetric scattering phantom and detect the 3D position of multiple embedded foreign bodies inside its 3D volume. Each foreign body (metallic spheres of $5\ \mathrm{mm}$ in diameter) is located  at a different depth inside the scattering phantom (dimensions $50\ \mathrm{mm}\times 50\ \mathrm{mm}\times 30\ \mathrm{mm}$). The phantom is illuminated from the position of the camera. The back-scattered computational pulses returning from each sphere indicate lateral as well as longitudinal position via its time of arrival. Figures~\ref{fig:Fig_4}c and d display two different ``slices'' at timestamps $28\ \mathrm{ps}$ and $56\ \mathrm{ps}$, which approximately correspond to depths of $2.8\ \mathrm{mm}$ and $5.6\ \mathrm{mm}$, respectively. Figure~\ref{fig:Fig_4}b shows a depth-resolved, volumetric reconstruction of all foreign bodies embedded within the scatterer.
We note that the object evaluation and  localization is performed automatically, without any prior knowledge of position or number of the embedded spheres. Instead, the mechanism exploits a fundamental property of SLiF that benefits the evaluation: In each acquired optical field measurement, the scattering in the phantom causes speckle. In moderate to strong scattering regimes, the signal back reflected from each small sphere becomes virtually indistinguishable from a speckle grain, if only one optical field at one wavelength would be captured. In contrast, SLiF acquires multiple optical fields that must remain correlated across the range of measured wavelengths for a signal to contribute to the synthetic wavelength reconstruction and subsequent pulse formation. Assuming sufficient SNR and the correct wavelength range, back-reflected signals from the embedded spheres, maintain this cross-wavelength speckle correlation, allowing constructive interference across synthetic wavelengths and producing a pulse. In contrast, the surrounding scattering medium introduces larger path length differences that lead to wavelength-dependent decorrelation, which contributes to speckle background suppression in the final SLiF reconstruction and provides a contrast improvement that allows to clearly separate the spheres from the background.

The full-field measurements demonstrated in this sub-section showcase SLiF's ability to not only provide 2D image content through scatterers, but also to time-gate the returning pulse with high resolution for a volumetric representation.  This ``slicing'' capability through scattering media at low instantaneous peak intensity hints at future potential use cases in applications which require precision measurements through difficult imaging conditions, such as biomedical imaging  through scattering tissue or virtual histology.

\vspace{-2mm}

\subsection{3D Imaging with Synthetic Pulses}
\label{subsec:3dimaging}

Similar to works in dual-wavelength interferometry \cite{fercher1985rough,dandliker1988two,zhou2022review}, Synthetic Wavelength Imaging has previously been used to effectively measure the depth of scenes and the 3D shape of rough surfaces \cite{willomitzer.2024, ballester.2024}. However, the use of a single synthetic field introduces a fundamental tradeoff between maximal axial measurement range and depth resolution: The maximal axial measurement range or ``unique measurement range'' is limited by $\Lambda/2$, meaning that larger synthetic wavelengths are needed to achieve larger unambiguous ranges. In contrast, smaller synthetic wavelengths are desired to achieve higher depth resolution. Similar to optical pulses in the optical domain, synthetic pulses resolve this tradeoff, i.e.,  simultaneously allow for large axial measurement ranges (limited by the pulse distance) and high depth resolution (limited by the width of each individual pulse).

\begin{figure}[t!]
\centering
\vspace{-6mm}
\includegraphics[width=\linewidth]{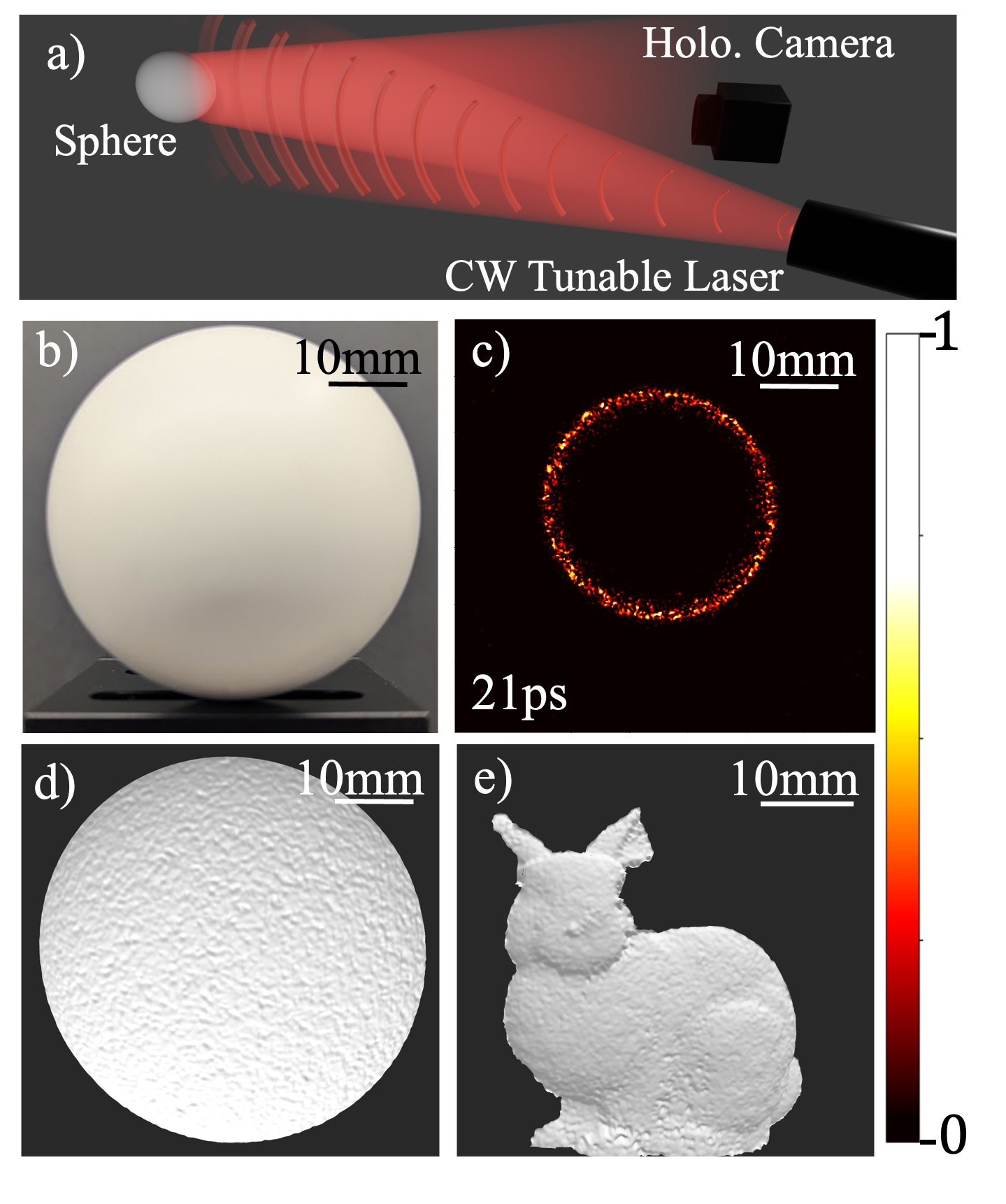}
\vspace{-10mm}
\caption{\textbf{ToF 3D Imaging of Optically Rough Surfaces using Band-Limited Synthetic Pulses:}
a) Setup schematic: The object (sphere) is illuminated by our band-limited CW tunable laser source and imaged with our holographic camera.
b) The measured object, a spray-painted ball bearing with $50.4\ \mathrm{mm}$ diameter.
c) Example synthetic pulse frame at $21\ \mathrm{ps}$.
d) 3D reconstruction of the object surface from the SLiF measurement at $2.41\ \mathrm{nm}$ laser bandwidth. The obtained depth precision is $340\ \mu\mathrm{m}$,  which is close to the theoretical axial resolution limit of $268\ \mu\mathrm{m}$.
e) Second example for 3D surfaces reconstruction: Bunny surface, reconstructed  from the measurements shown in Fig.~\ref{fig:Fig_1}e using $0.23\ \mathrm{nm}$ of bandwidth.}
\label{fig:Fig_5}
\end{figure}

To demonstrate the effectiveness of SLiF for full-field 3D imaging, we measured a spherical rough surface  ($50.4\ \mathrm{mm}$  diameter) using the full $2.41\ \mathrm{nm}$ (or $0.99\ \mathrm{THz}$) bandwidth of our lasers (see Fig.~\ref{fig:Fig_5}). 167 individual synthetic fields of the object (Fig.~\ref{fig:Fig_5}b) have been acquired. Figure~\ref{fig:Fig_5}c displays the resulting synthetic pulse-front, shown at a timestamp of $21\ \mathrm{ps}$. The 3D surface of the object was reconstructed by evaluating the timestamp (equivalent to path length) of the pulse maximum at any given pixel of the full-field SLiF measurement. The resulting 3D model is shown in Fig.~\ref{fig:Fig_5}d. After filtering out regions of low SNR caused by speckle-minima in the captured images, the depth precision (statistical variation of measured depth) of the measurement was evaluated by calculating the RMSE of the measured 3D point cloud w.r.t. a low frequency best fit  fourth-order polynomial surface. The obtained depth precision was $340\ \mu\mathrm{m}$,  which is close to the theoretical axial resolution limit of $268\ \mu\mathrm{m}$ that can be deduced for the used bandwidth in the absence of noise from literature in White Light Interferometry (WLI) or Optical Coherence Tomography (OCT) \cite{huang.1991, dresel.1992, Häusler.1998, wyant.2002}. A second evaluated 3D model  (bunny surface from Fig.~\ref{fig:Fig_1}) measured with 40 equally spaced  synthetic fields of $0.23\ \mathrm{nm}$ bandwidth, is shown in Fig.~\ref{fig:Fig_5}e.

The experiments shown in this section highlight again the relationship of our SLiF approach to techniques like WLI or OCT. In fact, SLiF provides the possibility to operate similar to a ``computational full-field OCT/WLI'' approach that is expected to reach the same competitive fundamental resolution limits, while the individual full-field capture of synthetic fields allows for greater flexibility and computational manipulation after the measurement, as discussed in the next section and Sec.~\ref{sec:discussion}. 

\subsection{Limitations and Tradeoffs}
\label{subsec:limitations}

While the previous theoretical considerations describe SLiF signal generation in the ideal noiseless case, real-world SLiF implementations are subject to acquisition noise and other hardware-level factors that can influence signal detection and resolution. Supplementary sec. 1 and 2 analyze both the ideal synthetic pulse and the impact of experimentally measured acquisition noise. By experimentally characterizing the noise in our optical field measurements and propagating it through the SLiF reconstruction pipeline, we have found good agreement between theory, simulation, and our shown experiments. It was further found that, for the noise levels observed in our system, the resulting phase noise primarily reduces pulse contrast by increasing the background floor and slightly decreasing the pulse peak, while causing no significant broadening of the synthetic pulse.

In addition to the discussion in sec. 2 of the supplementary material, the analysis presented in supplementary sec. 1 provides further insight into the impact of the spectral span $\Delta\lambda$ and the number of acquired optical fields $M$, on the temporal (or depth) resolution, unambiguous range, and acquisition time of SLiF measurements. It is shown that the total wavelength span (or, equivalently, optical frequency span) determines the achievable temporal pulse width, with larger bandwidths producing shorter synthesized pulses and therefore higher temporal resolution. In contrast, the spacing between adjacent wavelength samples determines the unambiguous temporal range. This results in an important trade-off: for a fixed bandwidth and uniformly spaced samples, reducing the wavelength spacing increases the unambiguous range but simultaneously increases the number of required measurements $M$, leading to longer acquisition times. While both the total bandwidth and wavelength spacing is adapted throughout this work to meet the requirements of individual experiments, most demonstrations employ $M=41$ closely spaced wavelength samples. This number was found to provide a practical balance between pulse quality, unambiguous range, and acquisition time, while still allowing the resulting SLiF datasets to be processed within minutes (depending on the desired temporal sampling and duration of the reconstructed video) using our current reconstruction algorithms.

For the acquisition time, the dominant bottleneck in our current implementation is the tuning speed of the employed laser source(s),  meaning  this limitation is not fundamental to the SLiF concept itself but rather reflects the characteristics of the current hardware implementation. The acquisition of a representative SLiF dataset consisting of $M=41$ optical fields presently requires approximately six minutes. In future demonstrations and potential subsequent real-world applications of SLiF, this bottleneck can be mitigated by exploiting multiple independent lasers at different center wavelengths, that either can be rapidly switched or can be combined with multiplexing  schemes similar to \cite{denz.1991, rubin.2017} that allow for the detection of even more optical fields simultaneously as previously shown in \cite{ballester.2024}. We note that neither long total acquisition times nor long camera exposure times inherently degrade the picosecond-scale temporal resolution achievable with SLiF. In fact, longer exposure times can be beneficial by improving the signal-to-noise ratio of the acquired optical fields. However, when imaging nominally static scenes that exhibit small residual motion, such as microscopic vibrations or mechanical drift, extended integration times may reduce fringe visibility through temporal averaging, leading to degraded field reconstruction quality.

A further tradeoff arises from the reconstruction procedure itself. Throughout this work, the amplitudes of all synthetic fields are normalized prior to pulse synthesis such that the reconstructed pulse is primarily determined by the recovered phase information. This choice reduces artifacts caused by spatial and spectral speckle fluctuations, which would otherwise introduce wavelength-dependent weighting during pulse synthesis. While amplitude normalization improves the homogeneity of the reconstructed pulses, phase estimates obtained from bright speckles are generally more reliable than those from dark speckles, suggesting that amplitude-aware pulse synthesis strategies represent an interesting direction for future research.

All these considerations place SLiF within a broader tradeoff space relative to state-of-the-art ultrafast imaging techniques, including SPAD-based imaging systems and Compressed Ultrafast Photography (CUP) \cite{gao.2014, gariepy.2015, liang.2018, morimoto2020megapixel, wang.2020, qi2020single, bocchieri2024scintillation}. To provide additional context, supplementary sec. 3 compares representative approaches \cite{gao.2014, morimoto2020megapixel} in terms of temporal resolution, spatial resolution, and robustness to non-cyclic scene dynamics, thereby highlighting the distinct operating regime occupied by SLiF.

In contrast to the temporal or depth resolution, which is theoretically bounded by the smallest \textit{synthetic} wavelength $\Lambda_{min} \propto 1/ \Delta\lambda_{max}$ used in the measurement, the optical lateral resolution remains bounded by the largest  \textit{optical} wavelength. This distinction arises because, in our approach, the acquired optical fields still serve as the carrier of the spatial information, while the synthetic wavelength determines the temporal/depth localization. Accordingly, in our shown  experiments, the effective spatial localization of the SLiF pulse field is primarily limited by the spatial resolution of the imaging system used to acquire and reconstruct the optical fields, i.e., the object-sided pixel pitch or the object-sided diffraction-limited resolution (whichever has a larger extent). Due to the requirements of our single-shot off-axis holography acquisition procedure \cite{ballester.2024}, the diffraction-limited lateral extend $\delta x$ is larger than the pixel pitch and hence constitutes the limiting factor in our system. $\delta x$ can be approximated by the Rayleigh criterion, $\delta x \approx \frac{1.22\,\lambda z}{D}$, where $\lambda$ is the respective optical wavelength, $z$ is the standoff distance and $D$ is the aperture diameter.  Given our current system parameters, this results in $\delta x \approx 85 \mu m$.

\vspace{-2mm}

\section{Synthetic Pulse Shaping}
\label{sec:pulseshaping}
One of the distinct advantages of our computational SLiF approach is the ability to manipulate pulses \textit{after} they have been acquired. Besides the basic phase alignment of all synthetic fields, discussed in sec.~\ref{sec:wavestopulses}, this unique property can be further expanded to \textit{freely shape the synthetic pulse in time and space}. From an information-theoretical perspective, temporal and spatial pulse shaping is equivalent to changing the information \textit{decoder} \cite{mait.2018, wagner.2003, hausler.2022}, and can be used to highlight or visualize different aspects or properties of the measured scene.

\vspace{-2mm}

\subsection{Temporal Pulse Shaping}
\label{subsec:temporal}
\begin{figure}[t!]
\centering
\vspace{-6mm}
\includegraphics[width=\linewidth]{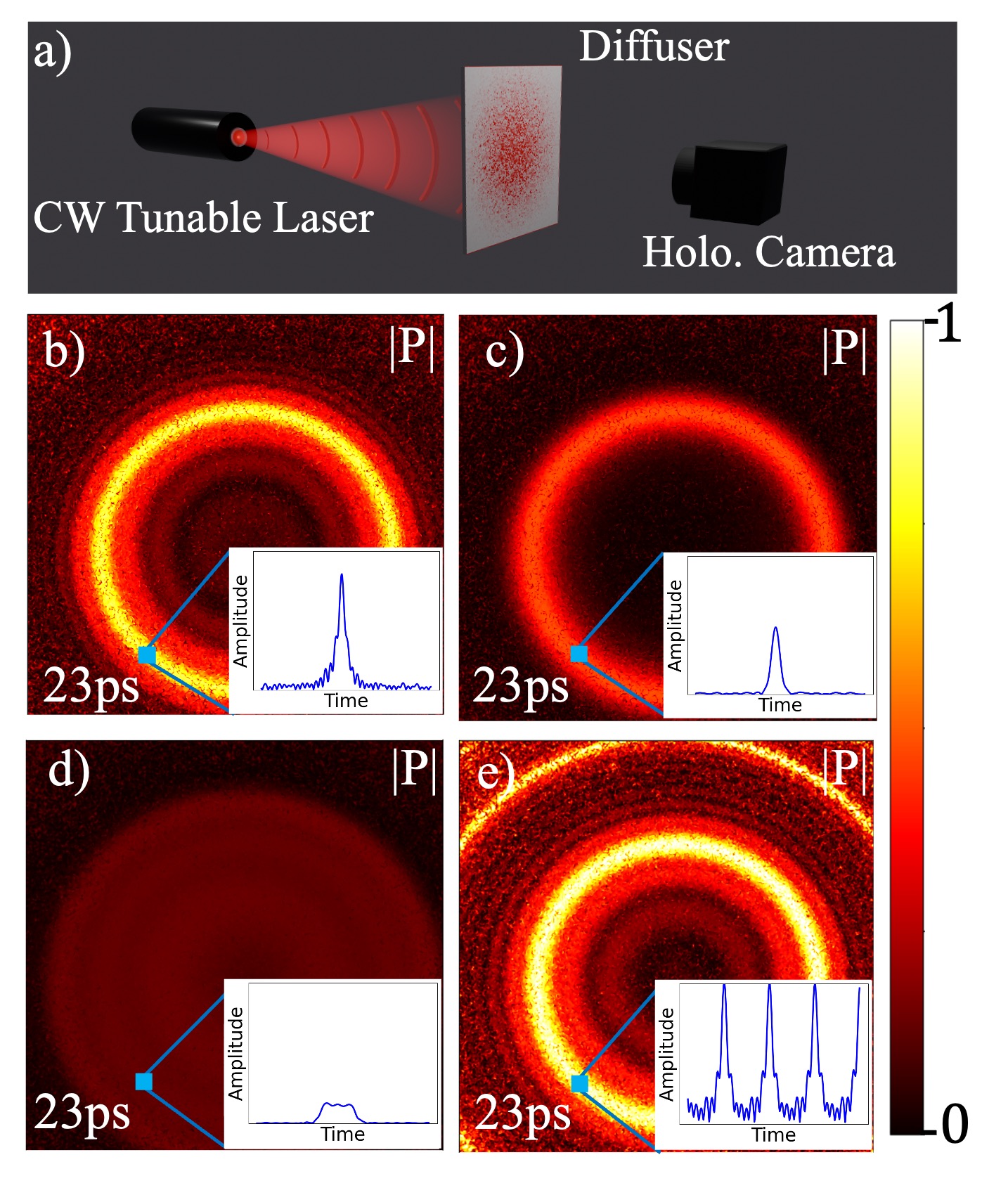}
\vspace{-10mm}
\caption{\textbf{Temporal Synthetic Pulse  Shaping (see videos \href{https://drive.google.com/drive/folders/1rS9Itz3QuB3RZksTBMV8XeLqz7HhctGi?usp=drive_link}{here} \cite{videos.2024}):}
a) Schematic of experimental setup: Point source laser illuminating a $75\times 75\ \mathrm{mm}$ diffuser in transmission
b)-d) Temporal pulse shaping of the spherical synthetic pulse-front incident on the diffuser. Windowing the amplitudes of the captured synthetic fields with different functions creates different temporal pulse shapes. Window functions: b) Rectangular, c) Hann window, d) $sinc$ function, e) Rectangular function using only 11 synthetic fields over an optical wavelength range between $854.20\ \mathrm{nm}$ and $856.05\ \mathrm{nm}$. Note that (b–e) plot $|P|$ instead of $|P|^2$ for better visibility of the temporal profiles.}
\label{fig:Fig_6}
\end{figure}

The experiments described in the last section involve the superposition of synthetic fields, where, according to Eq.~\ref{eq:addingfields}, each field is added with aligned phases and constant amplitudes. Due to the Fourier pair relationship between added fields and pulse shape, the resulting pulse train $P$ takes the form of a repeating $sinc$ function in the temporal domain. This can be observed in Fig.~\ref{fig:Fig_6}b: The experiment shows a spherical synthetic pulse-front incident on a planar ground glass diffuser. The light is emitted from a point source (fiber tip) centered behind the diffuser (Fig.~\ref{fig:Fig_6}a), which results in an imaged pulse-front in the spatial shape of an expanding circle (Fig.~\ref{fig:Fig_6}b see also video \href{https://drive.google.com/drive/folders/1rS9Itz3QuB3RZksTBMV8XeLqz7HhctGi?usp=drive_link}{here} \cite{videos.2024}). For the used optical wavelength range between $854.20\ \mathrm{nm}$ and $856.05\ \mathrm{nm}$, an experimentally evaluated FWHM of the synthetic pulse $|P|^2$ of $1.19\ \mathrm{ps}$ or $357\ \mu\mathrm{m}$ is achieved. Due to the high photon count regime of this measurement this is close to $355\ \mu\mathrm{m}$, which is the approximate theoretical linewidth for an intensity pulse created by rectangular sampling of $1.85\ \mathrm{nm}$  bandwidth.  

The basic idea of synthetic pulse shaping is motivated by the fact that each synthetic field used to create the pulse train can be  computationally manipulated individually, meaning that it is now possible to adjust the amplitude $\tilde{A}_{\Lambda_n}$ of each field to produce different temporal pulse shapes \textit{from the same set of measurements after their acquisition}. For example, adjusting the $\tilde{A}_{\Lambda_n}$ values to fit a Hann windowing function rather than a rectangle in the Fourier domain results in an apodized pulse shape (Fig.~\ref{fig:Fig_6}c, FWHM of $|P|^2$ is $2.07\ \mathrm{ps}$ or  $621\ \mu\mathrm{m}$). If a $sinc$ distribution is selected in the Fourier domain, a rectangular pulse shape will appear in the temporal domain (Fig.~\ref{fig:Fig_6}d). Moreover, it is possible to select a subset of the measured fields to adjust the length of the pulse train. By selecting fields with larger wavelength difference, the pulse train period is shortened so that a full period becomes observable in the given field of view (Fig.~\ref{fig:Fig_6}e). We note that Figs.~\ref{fig:Fig_6}b-e depict $|P|$ instead of $|P|^2$ for better visibility of the temporal profiles. Figures~\ref{fig:Fig_6}b-e  also show a plot of each temporal pulse profile, by selecting a single point in the rendered scene and graphing the amplitude in this point as a function of time (see video \href{https://drive.google.com/drive/folders/1rS9Itz3QuB3RZksTBMV8XeLqz7HhctGi?usp=drive_link}{here}\cite{videos.2024}). 

Besides the numerous potential advantages of computationally sharpening, broadening, or deforming the pulse peak after the measurements have been acquired, we envision several other potential future applications of temporal pulse shaping. For example, pulse pairs with a precisely defined distance could serve as a ``temporal ruler''  that could be flexibly moved back and forth in the computer to measure time or depth differences in the scene. In full-field mode, this could potentially be used as an ``adjustable virtual strobe illumination''  whose frequency could be selected very high, and dynamically changed after acquisition. Our pulse shaping approach could also be developed as a computational variant of earlier holographic pulse shaping approaches demonstrated, e.g., in \cite{hill.1993,hill.1995}. In another application scenario, a synthetic pulse sequence with different temporal shapes and intensity levels could be produced, which could potentially be used as a ``temporal code,'' similar to what has been proposed in \cite{gupta.2018} for amplitude-modulated time-of-flight cameras that work at much lower temporal and depth resolution.

\vspace{-1mm}

\subsection{Spatial Pulse Shaping}
\label{subsec:spatial}

\begin{figure}[b!]
\centering
\vspace{-6mm}
\includegraphics[width=\linewidth]{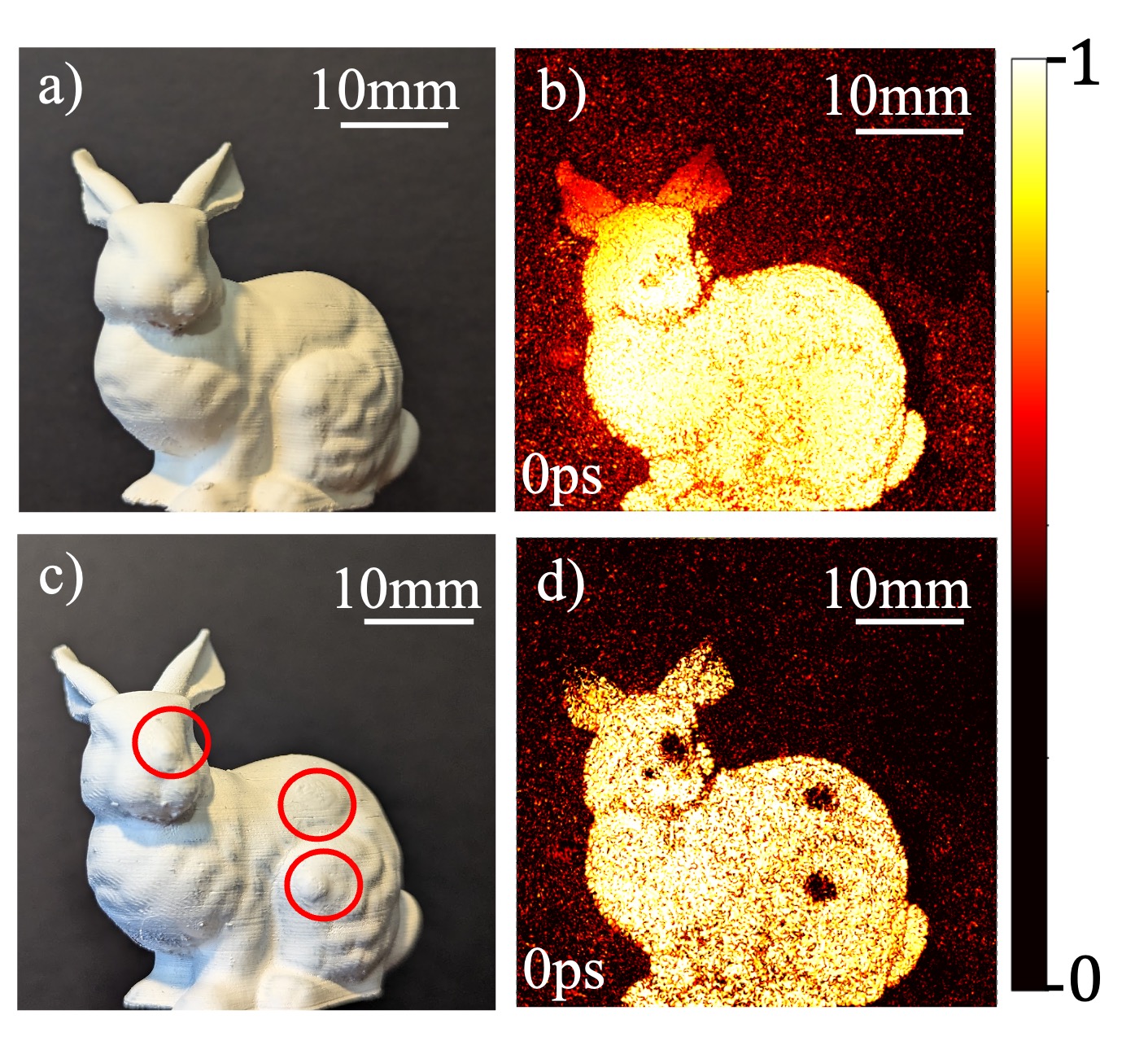}
\vspace{-10mm}
\caption{\textbf{Spatial Synthetic Pulse  Shaping and Evaluation of Surface Defects (see videos \href{https://drive.google.com/drive/folders/1rS9Itz3QuB3RZksTBMV8XeLqz7HhctGi?usp=drive_link}{here})\cite{videos.2024}:}
a) Image of the measured bunny object.
b) After spatial shaping of the spherical wavefront, the ``bunny-shaped'' pulse hits all surface points of a 3D-printed bunny object at the same time.
c) Small surface defects (paint droplets, less than $\sim$ $1\ \mathrm{mm}$ thick) are added to the object.
d) Using the same bunny-shaped wavefront to re-illuminate the altered object enables immediate visual identification of the defects without further image processing.}
\label{fig:Fig_7}
\end{figure}

In addition to the temporal pulse  shaping described above, a second type of computational pulse shaping is possible after measurement collection: spatial shaping. Spatial pulse shaping can be used to change the  3D spatial shape of the propagating synthetic pulse-front. An illustrative  example would be to turn a spherical pulse-front (as emitted from our point source-like fiber tip) into a planar pulse-front. We perform this spatial shaping by computationally delaying the synthetic pulse-front in each image pixel, where each scene point (pixel) receives a different delay so that a  certain pulse-front can be shaped. We first calibrate the initial 3D pulse shape produced by our setup by measuring the wavefront that is back scattered from a calibration object and incident on our camera sensor. Depending on the accuracy requirements, the wavefront shape can be obtained from a full SLiF pulse or just a single captured synthetic field. The calibration objet can be a planar surface similar to Fig.~\ref{fig:Fig_6}b or c, or a freeform object with known shape.  Eventually, we computationally add a custom phase delay to every pixel $(x,y)$ of each synthetic field $E(\Lambda_n)$. This phase delay is calculated from the difference between the calibrated 3D pulse shape and the ``target 3D shape''  of the desired pulse-front. We emphasize that this ``target 3D shape''  could be a mathematical function (plane, sphere, paraboloid, etc.) or a freeform 3D shape (bunny, face, technical part, etc.). The 3D shape can be imported from a CAD-file, measured by an independent 3D sensor, or even measured with the SLiF 3D sensing capabilities introduced in sec.~\ref{subsec:3dimaging}.

Figure~\ref{fig:Fig_7} visualizes the process of spatial pulse shaping and a potential application: After taking a SLiF measurement of a scene with a printed 3D object (bunny, Fig.~\ref{fig:Fig_7}a) and performing the steps mentioned above, the object is virtually illuminated with a synthetic pulse-front that has been spatially reshaped to the 3D shape of a bunny. As all points in the bunny-shaped pulse-front hit the surface of the 3D object at the same time, all surface points ``flash up''  simultaneously for one timestamp (arbitrarily set to $0\ \mathrm{ps}$ in Fig.~\ref{fig:Fig_7}b). In the following step, we add small ``defects'' to the surface by applying small white paint droplets. An image of the bunny figure ``with defects''  is shown in Fig.~\ref{fig:Fig_7}c. Each defect is only approximately $3\ \mathrm{mm}$ wide and $1\ \mathrm{mm}$ thick, and spotting the defects in the conventional camera image of Fig.~\ref{fig:Fig_7}c is challenging. However, when illuminating the 3D object (with defects) with the previously created bunny-shaped wavefront (without defects), the three applied defects are immediately visible in the image produced for the $0\ \mathrm{ps}$ timestamp (Fig.~\ref{fig:Fig_7}d), as the pulse-front now differs from the true 3D shape at the positions of the defects. We believe that this measurement highlights the great potential of our method for defect detection in industrial inspection, but also for the detection of irregularities in medical images, forensic scenes, or works of art. An image like Fig.~\ref{fig:Fig_7}d can be easily processed with basic image processing algorithms and fed into existing pipelines, e.g., in an industrial  production line.
\vspace{-2mm}

\section{Material Properties}
\label{sec:materialproperties}

\begin{figure}[t!]
\centering
\vspace{-6mm}
\includegraphics[width=\linewidth]{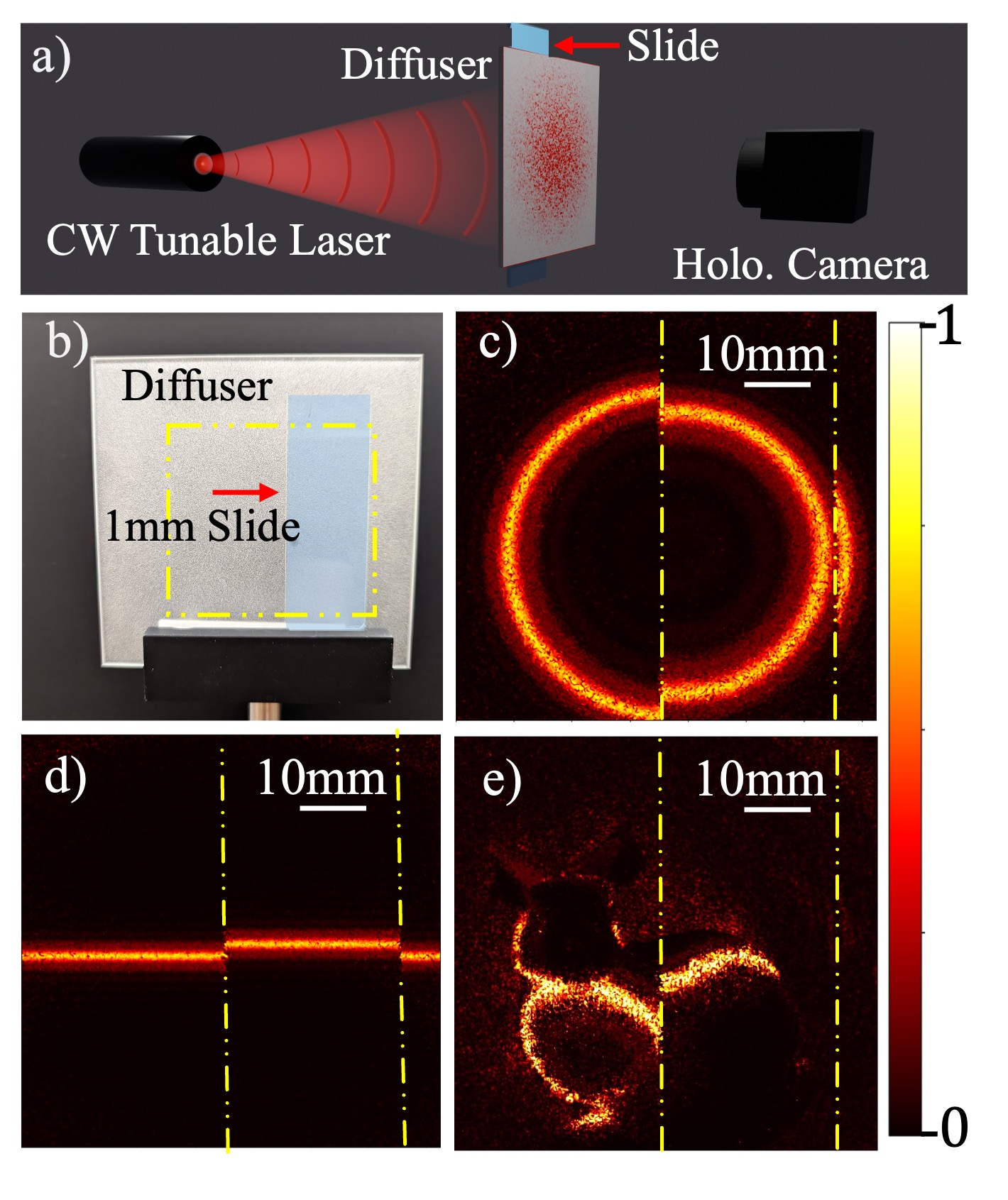}
\vspace{-10mm}
\caption{\textbf{Evaluation of Material Properties and Decoding via Synthetic Pulse Shaping (see videos \href{https://drive.google.com/drive/folders/1rS9Itz3QuB3RZksTBMV8XeLqz7HhctGi?usp=drive_link}{here})\cite{videos.2024}:}
a) Schematic of measurement: A point light source illuminating (in transmission) a $75\times 75\ \mathrm{mm}$ diffuser as well as a $1\ \mathrm{mm}$ thick glass slide covering a portion of the field of view.
b) Picture of the diffuser and slide (highlighted in blue) inside the camera's field of view (yellow dashed line).
c)~Synthetic pulse, computed as in Fig.~\ref{fig:Fig_6}b. The right part of the pulse is delayed by the optical path length difference introduced by the glass slide.
d) Synthetic pulse shaping allows for the manipulation of the pulse-front to create additional forms of decoding. Here, a ``scan line'' is created from the same set of measurements. Pulse delay due to glass slide visible in the line offset.
e) A bunny-shaped wavefront is created from the same set of measurements. The pulse-offset is also visible in the bunny-shaped wavefront.}
\label{fig:Fig_8}
\end{figure}

Although the pulses presented in this paper do not physically exist and only exist in the computer, they are still subject to the physical constraints of the scene. An example of this can be shown by observing the material properties of the objects in the scene, including their refractive indices. Exactly as optical pulses, synthetic pulses slow down when passing through materials with higher refractive indices. The experiments shown in Fig.~\ref{fig:Fig_8} demonstrate this effect:  Similar to Fig.~\ref{fig:Fig_6}, a diffuser is illuminated with a point source from behind, but now an additional  $\sim$ $1\ \mathrm{mm}$ thick glass slide covers a part of the diffuser (see Fig.~\ref{fig:Fig_8}a). The evaluated SLiF video (see Fig.~\ref{fig:Fig_8}c and full video \href{https://drive.google.com/drive/folders/1rS9Itz3QuB3RZksTBMV8XeLqz7HhctGi?usp=drive_link}{here}~\cite{videos.2024}) shows that the synthetic pulse-front traveling through both diffuser and glass slide is clearly delayed with respect to the pulse-front that only travels through the diffuser, which is caused by the increased optical path length introduced by the glass. After measuring the glass thickness to $1.06\ \mathrm{mm}$, we evaluated the refractive index of the glass to 1.54, which is a common value for glass slides. Figures~\ref{fig:Fig_8}d~and~e show the same delay effect, but demonstrate additional forms to decode the increased optical path length with synthetic pulse shaping. In Fig.~\ref{fig:Fig_8}d, the original spherical pulse-front is computationally reshaped to a planar pulse-front with a tilt in the vertical direction. When propagating through the flat diffuser with glass slide, the pulse-front creates a horizontal ``scan line''  which experiences a vertical offset from the increased optical path length in the glass. In Fig.~\ref{fig:Fig_8}e the diffuser is illuminated with a bunny-shaped wavefront, similar to Fig.~\ref{fig:Fig_7}.  While this experiment is meant to emphasize the flexibility of the approach (i.e., arbitrary spatial pulse shapes are possible), a clear pulse-front offset from the glass slide is still visible in the bunny profile. As before, all pulse fronts have been created from the same set of measurements after the measurements have been taken.

Although not shown here for the sake of brevity, \textit{temporal} pulse shaping could also be exploited to further decode potentially important information in this scene. For example, a ``double-peak''  pulse-front could be engineered where the temporal difference between both peaks exactly lines up with the optical path length difference introduced by the glass slide. For a pulse with a planar-shaped front parallel to the diffuser surface, this would mean that the glass slide and diffuser surface flash up at the exact same timestamp. This procedure could be used to detect thickness or refractive index variations in the glass slide. We believe that quantifying material properties with synthetic pulses has important potential application in the characterization of optical components (e.g., testing of multi-surface lens systems, optical flats or mirrors), or the characterization of thin film materials~\cite{bass.2024}. Moreover, the ``sectioning''  ability of synthetic pulses paired with their robustness to scattering could potentially lead to impactful applications in medical imaging, such as novel pump-and-probe methods or virtual, optical stain-free histology which would not require slicing the sample and could potentially even be performed in vivo.

\section{Discussion}
\label{sec:discussion}

We presented a novel method that utilizes holographic information acquired at several closely spaced optical wavelengths to generate synthetic pulses from multiple synthetic fields, enabling the visualization of light as it propagates through arbitrary scenes. Our measurement scheme leverages the holographic nature of the data to acquire full-field light-in-flight videos in a large FoV using tunable CW lasers and conventional CMOS focal plane arrays, thereby eliminating the need for pulsed lasers or costly high-speed detectors. As demonstrated in the previous sections, we recover LiF videos of a wide range of scenes, including volumetric 3D environments, scattering media, and materials with spatially varying refractive indices, with a temporal resolution of around $1 \ \mathrm{ps}$ (see measurements  performed at $\Delta \lambda = 2.41$~$\mathrm{nm}$ in Fig.~\ref{fig:Fig_5} and $\Delta \lambda = 1.85$~$\mathrm{nm}$ in Fig.~\ref{fig:Fig_6}). As discussed in sec.~\ref{subsec:limitations}, this resolution is currently limited by the spectral tuning range of our lasers and can be readily improved by employing sources with broader tuning bandwidths. In addition, we demonstrated the unique strengths of our approach, particularly its flexibility and reconfigurability, by spatio-temporally shaping the synthetic pulse post-acquisition for diverse potential applications such as defect detection and material characterization.

Our demonstrated system can emulate a wide range of pulsed illumination patterns from a single set of measurements, with the temporal and spatial structure of the effective illumination field being freely tailored after acquisition. The flexibility of this approach becomes particularly apparent when compared to conventional pulse-generation hardware, which could, in principle, realize similar pulse-shaping tasks but typically requires substantial hardware modifications whenever the pulse structure should be changed. For example, altering the repetition rate or temporal profile of a mode-locked laser may require adjustments to the cavity length, the addition of pulse pickers, modifications to synchronization electronics, or dedicated pulse-shaping hardware. Such constraints make it challenging to rapidly explore arbitrary temporal illumination patterns, especially when simultaneous spatial shaping (e.g., a ``bunny-shaped'' pulse front) is desired. In contrast, SLiF enables both spatial and temporal pulse shaping to be performed computationally from the same acquired dataset, without requiring changes to the illumination or detection hardware.

Another potential advantage of SLiF arises from the fact that the recovered pulse-like temporal information is generated solely from continuous-wave illumination. While the reconstructed light field exhibits pulse-like propagation and time-of-flight behavior, the physical probing field remains a low-peak-power CW field throughout the measurement. This creates a measurement regime that differs fundamentally from conventional light-in-flight approaches based on ultrashort optical pulses. In such systems, increasing the signal level can be limited by peak-intensity-driven effects including Kerr nonlinearities, multiphoton absorption or ionization, Raman scattering, and associated changes in coherence \cite{agrawal.2000,boyd.2008}. These effects can alter the propagation dynamics, local refractive index, or scattering behavior of the system under investigation, making the measurement itself partially invasive. By contrast, SLiF decouples the recovery of pulse-like temporal information from the use of high instantaneous optical power, potentially enabling time-resolved measurements under conditions that remain closer to the natural operating regime of the device or material being studied.

One application area where this distinction may become particularly relevant is the characterization of nonlinear photonic devices, optical fibers \cite{balaji.2026}, integrated photonic circuits, and resonant photonic structures such as high-Q ring resonators. In these systems, pulsed excitation can induce intensity-dependent refractive-index changes, resonance detuning, nonlinear scattering processes, or coherence degradation, potentially obscuring weak defect-induced signatures or modifying the very propagation behavior that one aims to measure \cite{vahala.2003}. A SLiF measurement, in contrast, can probe such devices using low instantaneous optical intensity while still providing pulse-like, temporally resolved information about propagation, delay, and scattering. In this application regime, we therefore envision SLiF not merely as an alternative implementation of time-resolved imaging, but as a complementary measurement modality that may offer unique advantages when characterizing photonic systems under operating conditions that are difficult to access with conventional pulsed approaches.

Throughout this work, we restricted our discussion to optical fields acquired at uniformly spaced wavelengths. However, future implementations of SLiF may benefit from irregular or nonuniform wavelength sampling strategies. Similar to approaches explored in enhanced-resolution imaging, nonuniform sampling theory, and compressed sensing, appropriately chosen wavelength distributions can provide additional flexibility in balancing temporal resolution, unambiguous range, and acquisition time. When combined with reconstruction algorithms that accurately incorporate the underlying forward model, such sampling strategies may enable improved recovery of depth and temporal features beyond what would be expected from a simple uniform-sampling interpretation. A more detailed discussion of these possibilities is provided in supplementary sec. 1B.

As discussed, our basic SLiF approach has drawn inspiration from earlier interferometric and holographic LiF techniques \cite{arons.1995, Häusler.1996, inoue.2023, shih.1999, marron.1992}. We further note that our demonstrated spatio-temporal pulse shaping capabilities bear certain conceptual similarities to recent ToF approaches, such as \cite{sultan.2024}, where, the measured ToF information is manipulated after acquisition through the application of phase or temporal delays. These delays can be chosen such that the pulse fronts can subsequently be focused, shaped, and visualized. However, the underlying measurement principles are fundamentally different. The above approaches operate on transient measurements acquired using pulsed illumination and ultrafast detectors and often require the acquisition of large transport datasets, sometimes involving scanning procedures or comparatively sparse spatial sampling, and are also limited by the limited temporal resolution of the electronic detectors. In contrast, SLiF directly operates on dense full-field optical measurements and therefore combines interferometric field acquisition with computational pulse synthesis and shaping capabilities. Beyond these conceptual parallels in computational pulse synthesis, some of the most relevant comparisons for SLiF can be found among coherence-gating and related interferometric ranging techniques, such as optical coherence tomography (OCT) ~\cite{huang.1991, Häusler.1998}, coherence radar, and white-light interferometry (WLI) ~\cite{dresel.1992, deck.1994}. In fact, SLiF can be seen as a \textit{``computational OCT/WLI advancement''} that exhibits several important distinctions from the ``classical'' OCT/WLI techniques. A key difference is that each synthetic field is acquired independently instead of integrating the detection over a broadband or chirped signal, or moving a mirror in the reference arm of a short-coherence interferometer. This results in a fairly long acquisition time needed to capture the large number of synthetic fields to construct a single SLiF measurement. Although this inherent drawback is still a bottleneck of our current method, it may be mitigated in the future through clever multiplexing strategies, several of which have already been explored in the holography literature (see our discussion on sec. \ref{subsec:limitations}). Another difference and technical constraint of our approach is that the current spectral bandwidth of our prototype system is limited, which bounds our minimal achievable synthetic pulse width. 

However, the property that each synthetic field is acquired separately also results in important benefits of our method, which distinguishes SLiF from current OCT or WLI systems. In contrast to many common OCT systems,  our  measurements are always performed in a full-field configuration, without the need for raster scanning or B-scans. When imaging through scattering media, the captured synthetic fields typically contain most of the scattered photons. This means that, in contrast to OCT,  our SLiF approach does not rely on confocal gating and has the potential to image deeper inside the scatterer (fog, tissue, etc).

Compared to a related approach that  captures and sums up \textit{optical} fields \cite{arons.1995}, the superposition of synthetic fields has distinct advantages: Synthetic fields are less prone to decorrelation if the object or the scatterer that embeds the object moves, which allows computational pulse generation from many sequentially captured fields through moving scatterers, such as living tissue or fog. Moreover, in contrast to optical fields, synthetic fields are largely speckle-free, meaning that their phase information can be used to computationally compensate for laser drift, as shown in this paper.  This property  allows our SLiF method in the end to computationally shape the pulses in time and space after their acquisition, which emphasizes again the computational character of our method compared to classical OCT/WLI approaches.

In the future, we hope that the capabilities introduced by SLiF can be further used to advance multiple fields in computational imaging and potentially lead to a new class of cameras for visualizing and quantifying physical processes in various applications such as industrial inspection, photonic device characterization, automotive sensing, or biomedical imaging in deep tissue. \\

\section*{Declarations}
\subsection*{Availability of data and materials}
The datasets used and/or analyzed during the current study are available from the corresponding authors on reasonable request.
\subsection*{Competing interests}
The authors declare that they have no competing interests.
\subsection*{Funding}
This project was partially funded by the Optica Foundation 20th Anniversary Challenge Award.
\subsection*{Authors' contribution}
F.W secured funding. P.C. and F.W. conceived of ideas. M.B., P.C. and F.W. built the hardware and software used during experimentation. P.C. and S.F. conducted the experiments. P.C., M.M.B., F.W., and A.K. disscussed and evaluated the experimental results. All authors contributed to the preparation of the manuscript. 
\subsection*{Acknowledgments}
This work was supported by [funding information will be included in final manuscript]. We acknowledge Heming Wang and Parker Liu for their help in hardware design and assistance in code writing. We also thank R. Jason Jones for the fruitful discussions about comparisons with existing pulse based methods. Moreover, we thank Gerd Häusler for providing comments and feedback on the initial manuscript draft. A preprint version of this manuscript was submitted to arXiv on July 10, 2024 \cite{cornwall.2024}.


\section*{References}

\bibliography{MyLibrary}

\section*{Keywords}
Digital Holography, Interferometry, Synthetic wavelength, Light-in-flight, Multiwavelength interferometry, Scattering

\section*{List of abbrevations}
CW- Continuous Wave \\
FWHM- Full-Width-at-Half-Maximum \\ 
LiF- Light-in-flight \\
OCT- Optical Coherence Tomography\\ 
SLiF- Synthetic Light-in-Flight\\
SPAD- Single-Photon Avalanche Diode\\ 
SWI- Synthetic Wavelength Imaging\\ 
WLI- White Light Interferometry\\ 

\end{document}